\documentclass[12pt,aps,prd,showpacs,amsmath,amssymb]{revtex4}
\usepackage{graphicx}
\usepackage{multirow}
\usepackage{subfigure}
\input epsf
\begin{document}
\title{The temperature dependence of the decuplet baryon masses from thermal QCD sum rules}
\author{Yong-Jiang Xu, Yong-Lu Liu\footnote{yongluliu@nudt.edu.cn}, and Ming-Qiu Huang}
\affiliation{ College of Science, National University of Defense Technology, Hunan 410073, China}
\date{\today}
\begin{abstract}
In the present work, the masses of the decuplet baryons at finite temperature are investigated using thermal QCD sum rules. Making use of the quark propagator at finite temperature, we calculate the spectral functions to $T^{8}$ order, and find that there are no contributions to the spectral functions at $T^{8}$ order and the temperature corrections mainly come from that containing $T^4$ ones. The calculations show very little temperature dependence of the masses below $T=0.11\mbox{GeV}$. While above that value, the masses decrease with increasing temperature. The results indicate that the hadron-quark phase transition temperature may be $T_c\geq0.11\mbox{GeV}$ for the decuplet bayons.
\end{abstract}
 \maketitle

\section{Introduction}\label{sec1}

The study of hadronic properties under extreme conditions of temperature and density is an interesting and important theoretical thesis, being related to present and future heavy-ion experiments. Whereas the calculation of hadronic parameters at finite temperature and density directly from QCD faces some difficulties. The thermal QCD is a calculable theory in large momentum transfer region or at short distance, where the running coupling constant is small, so the perturbative method can be used efficiently. However, the coupling constant becomes large and the standard perturbation theories fail at the hadronic level. Therefore, investigation of hadron properties requires nonperturbative methods. Some nonperturbative approaches have been put forward since 1970s, such as lattice QCD, heavy quark effective theory (HQET), different quark models, QCD sum rules and so on. In addition, it is believed that the hadronic matter undergoes a phase transition to a quark-gluon plasma(QGP) at sufficiently high temperature. Whereas the nature of this transition is not well understood, so investigation of thermal properties of hadrons is very instructive.
\par The QCD sum rule method has been widely used to study the thermal properties of the light-light mesons in Refs. \cite{art1,art2}, the heavy-light mesons in Refs. \cite{art3,art4,art5} and the heavy-heavy mesons in Refs. \cite{art6,art7,art8,art9,art10,art11} since it was firstly extended to finite temperature in Ref. \cite{art12} by Bochkarev and Shaposhnikov. The extension bases on the following assumption: Both the operator product expansion(OPE) technique and notion of quark-hadron duality remain valid at finite temperature, but the vacuum condensates must be replace by their thermal expectation values. For baryons, Yuji. Koike investigated the octet baryon masses at finite temperature, taking into account the contribution of $\pi$+$N$ $\rightarrow N$ in the calculation of the spectral function\cite{art13}. In Ref. \cite{art14}, the authors constructed the spectral representation of the two-point correlation function of nucleon from the Feynman diagrams. Therefore it is interesting to investigate the temperature effect of the baryon masses from the principle of QCD. The main aim of this work is to calculate the spectral functions directly from the quark propagator at finite temperature.
\par The rest of the paper is organized as follow. In section \ref{sec2}, we calculate the spectral functions to $T^{8}$ order, summarize the nonperturbative contributions and obtain the Borel transformed sum rules for the masses. Section \ref{sec3} is devoted to the numerical analysis and a summary is given at the end of this part.

\section{Thermal QCD sum rules of the decuplet baryons}\label{sec2}

 We begin with the following two-point thermal correlation function:
 \begin{equation}\label{correlator}
 \Pi_{\mu\nu}(q,T)=i\int dx^{4}e^{iqx}\langle \textsl{T}(J_{\mu}(x))\overline{J_{\nu}(0)}\rangle,
 \end{equation}
 where $J_{\mu}(x)$ is the interpolating current for the decuplet baryons. In the above equation, the first $T$ is a variable for temperature and the second one stands for time-order product. We omit the first one in the rest of the paper in order to avoid confusion. The thermal average of any operator $O$ is defined as
 \begin{equation}
 \langle O \rangle=\frac{Tr(\exp(-\beta H)O)}{Tr(\exp(-\beta H))},
 \end{equation}
 in which $H$ is the QCD Hamiltonian, $\beta=\frac{1}{T}$ stands for the inverse of the temperature, and the traces are carried out over the complete set of states. In this paper, we use the Ioffe's currents for the baryons:
 \begin{eqnarray}
 &&J_{\mu}^{\Delta}=\varepsilon_{abc}(u^{a}(x)C\gamma_{\mu}u^{b}(x))u^{c}(x),\nonumber\\
 &&J_{\mu}^{\Sigma^{*}}=\sqrt{\frac{1}{3}}\varepsilon_{abc}({2(u^{a}(x)C\gamma_{\mu}s^{b}(x))u^{c}(x)+(u^{a}(x)C\gamma_{\mu}u^{b}(x))s^{c}(x)}),\nonumber\\
 &&J_{\mu}^{\Xi^{*}}=\sqrt{\frac{1}{3}}\varepsilon_{abc}({2(s^{a}(x)C\gamma_{\mu}u^{b}(x))s^{c}(x)+(s^{a}(x)C\gamma_{\mu}s^{b}(x))u^{c}(x)}),\nonumber\\
 &&J_{\mu}^{\Omega}=\varepsilon_{abc}(s^{a}(x)C\gamma_{\mu}s^{b}(x))s^{c}(x),
 \end{eqnarray}
 where $a,b,c$ are color indices and $C$ is the charge conjugate operator.
\par  According to the standard procedure of the QCD sum rules, we need to calculate the correlator(\ref{correlator}) in terms of physical particles (hadrons) and in quark-gluon language, and then match the two representations. On the theoretical side, the correlator can be expanded as
 \begin{equation}
 \Pi(q^2)=\Pi^{pert}(q^2)+\Pi^{cond}(q^2)=\int ds\frac{\rho(s)}{s-q^{2}}+\Pi^{cond}(q^2),
 \end{equation}
 where $\rho(s)$ is the spectral density. We calculate the perturbative part of the two-point thermal correlator by using the quark propagator at finite temperature in coordinate space\cite{art15},
 \begin{equation}
 S_{T}(x)=S_{0}(x)f(\pi T x),
 \end{equation}
 where $S_{0}(x)$ is the quark propagator at $T=0$ and $f$ is a temperature modification factor. The function $f(\pi T x)$ satisfies $f(0)=1$ at $T=0$ and at $T\neq 0$
 \begin{equation}
 f(z)=z \exp(-z) \frac{z+1+(z-1)\exp(-2z)}{[1-\exp(-2z)]^{2}},
 \end{equation}
 which can be expanded at small $z$ as
 \begin{equation}
 f(z)=1-\frac{7 z^{4}}{360}.
 \end{equation}
 For $u$ or $d$ quark, $S_{0}(x)=\frac{i\not\!{x}}{2 \pi^{2} x^{4}}$, and for $s$ quark, $S_{0}(x)=\frac{i\not\!{x}}{2 \pi^{2} x^{4}}-\frac{m}{4 \pi^2 x^2}$, where $m$ is the mass of the $s$ quark.
\par After some calculations, it is found that the terms containing $T^{8}$ do not contribute to the imaginary part of the perturbative side of the two-point thermal correlator. The result for the $\Delta$ is
\begin{eqnarray}\label{imaginary}
\mbox{Im}\Pi_{\mu\nu}(q)=&&(-\frac{1}{20 (2\pi)^{3}} q^{4}-\frac{14}{15(2\pi)^{3}}\pi^{4} T^{4})g_{\mu\nu}\not\!{q}
                  +\frac{1}{20 (2\pi)^{3}} q^{2}q_{\mu}q_{\nu}\not\!{q}+\nonumber\\ &&\frac{1}{64 (2\pi)^{3}} q^{4}\gamma_{\mu}\gamma_{\nu}\not\!{q}
                  -\frac{7}{20 (2\pi)^{3}}\pi^{4} T^{4}\gamma_{\mu}\gamma_{\nu}\not\!{q}
                  +(\frac{1}{80 (2\pi)^{3}} q^{4}-\nonumber\\ &&\frac{7}{15 (2\pi)^{3}} \pi^{4} T^{4})\gamma_{\nu} q_{\mu}+(-\frac{3}{160 (2\pi)^{3}} q^{4}+\frac{7}{30 (2\pi)^{3}} \pi^{4} T^{4}) \gamma_{\mu} q_{\nu}.
\end{eqnarray}
 According to the basic assumption of the extension of the QCD sum rules to finite temperature, we take advantage of the results in Ref. \cite{art16} for the condensate part, replacing the vacuum condensates by the thermal expectation values,
 \begin{eqnarray}\label{condensate}
 \Pi_{\mu\nu}^{cond}(q)=&&-\frac{1}{3 \pi^{2}} \langle\overline{u} u\rangle q^{2} \ln(-q^{2})\{g_{\mu\nu}
                        -\frac{5}{16} \gamma_{\mu}\gamma_{\nu}+\frac{1}{4}(\gamma_{\mu} q_{\nu}-\gamma_{\nu} q_{\mu})\frac{\not\!{q}}{q^{2}}
                        -\frac{1}{2}\frac{q_{\mu}q_{\nu}}{q^{2}}\}\nonumber\\
                        &&+\frac{4}{3}\langle\overline{u}u\rangle^{2}\frac{1}{q^{2}} \{g_{\mu\nu}\not\!{q}
                        -\frac{3}{8} \gamma_{\mu}\gamma_{\nu}\not\!{q}
                        +\frac{3}{8} (\gamma_{\mu} q_{\nu}-\gamma_{\nu}q_{\mu})
                        -\frac{1}{8} (\gamma_{\mu}q_{\nu}+\gamma_{\nu}q_{\mu})\}\nonumber\\
                        &&+\frac{1}{24\pi^{2}}\ln(-q^{2})g_{s}
                        \langle\overline{u}\sigma_{\lambda\sigma}\frac{\lambda^{a}}{2}u G_{\lambda\sigma}^{a}\rangle(g_{\mu\nu}-\frac{1}{4}\lambda_{\mu}\lambda_{\nu}).
 \end{eqnarray}
\par Now we turn to the hadronic representation of the thermal two-point correlator(\ref{correlator}). In order to obtain the hadronic representation, a complete set of states $\Delta(q,r)$ are inserted into the r.h.s. of Eq. (\ref{correlator}),
\begin{equation}
\langle T(J_{\mu}(x))\overline{J_{\nu}(0)}\rangle=\sum_{r}\langle T(J_{\mu}(x))|\Delta(q,r)\rangle\langle\Delta(q,r)|\overline{J_{\nu}(0)}\rangle,
\end{equation}
where the summation is made over the spin projection $r$ of the baryon $\Delta$ with momentum $q$, and $q^2=M_{\Delta}^2$, with $M_{\Delta}$ the mass of $\Delta$. The coupling of the interpolating current to the baryon state is defined as
\begin{equation}
\langle 0|J_{\mu}(0)|\Delta(q,r)\rangle=\lambda_{\Delta} \nu_{\mu}^{r}(q),
\end{equation}
in which $\nu_{\mu}^{r}(q)$ is the wave function of the $\Delta$ baryon, $(\not\!{q}-M_{\Delta})\nu_{\mu}^{r}(q)=0$, and $\lambda_{\Delta}$ is a constant. Then the hadronic representation of the correlator is obtained as
\begin{eqnarray}\label{hadronic}
&&\sum_{r}\langle (J_{\mu}(x))|\Delta(q,r)\rangle\langle\Delta(q,r)|\overline{J_{\nu}(0)}\rangle\nonumber\\ &&=-\lambda_{\Delta} [g_{\mu\nu}\not\!{q}-\frac{1}{3}\gamma_{\mu}\gamma_{\nu}\not\!{q}+\frac{1}{3}(\gamma_{\mu}q_{\nu}-\gamma_{\nu}q_{\mu})-\frac{2}{3}\frac{q_{\mu}q_{\nu}\not\!{q}}{M_{\Delta}^{2}}\nonumber\\ &&+g_{\mu\nu}M_{\Delta}-\frac{1}{3}\gamma_{\mu}\gamma_{\nu}M_{\Delta}+\frac{1}{3}(\gamma_{\mu}q_{\nu}-\gamma_{\nu}q_{\mu})\frac{\not\!{q}}{M_{\Delta}}-\frac{2}{3}\frac{q_{\mu}q_{\nu}}{M_{\Delta}}]+\cdots,
\end{eqnarray}
where $``\cdots"$ stands for the contributions of continuum states and the interaction between the current and  particles in the medium\cite{art16}. It can be seen that the structures $g_{\mu\nu}$ and $g_{\mu\nu}\not\!{q}$ contain more information than other structures by comparing the Eqs.(\ref{imaginary})-(\ref{condensate}) with Eq. (\ref{hadronic}), therefore we chose the two structures, $g_{\mu\nu}$ and $g_{\mu\nu}\not\!{q}$, to obtain the sum rule. For the structure $g_{\mu\nu}$, the spectral function is
\begin{equation}
\rho_{1}(s)=\lambda_{\Delta}^{2}M_{\Delta} \delta(s-M_{\Delta}^{2})+\cdots,
\end{equation}
and for the structure $g_{\mu\nu}\not\!{q}$, the spectral representation is
\begin{equation}
\rho_{2}(s)=\lambda_{\Delta}^{2} \delta(s-M_{\Delta}^{2})+\cdots.
\end{equation}
According to the quark-hadron duality, the contributions of the continuum states and the interaction between  current and  particles in the medium can be approximated by the OPE spectral function. Finally we arrive at
 \begin{equation}
\Pi_{1}^{phy}=\int ds\frac{\rho_{1}(s)}{s-q^{2}}=\frac{\lambda_{\Delta}^2 M_{\Delta}}{M_{\Delta}^2-q^{2}}+\int_{s_{0}}^{\infty} ds \frac{\frac{-1}{3 \pi^{2}} \langle\overline{u} u\rangle s}{s-q^2},
\end{equation}
\begin{equation}
\Pi_{2}^{phy}=\int ds\frac{\rho_{2}(s)}{s-q^{2}}=\frac{\lambda_{\Delta}^2} {M_{\Delta}^2-q^{2}}+\frac{1}{\pi}\int_{s_{0}}^{\infty} ds \frac{-\frac{1}{20 (2\pi)^{3}} s^{2}-\frac{14}{15(2\pi)^{3}}\pi^{4} T^{4}}{s-q^{2}},
 \end{equation}
 with $s_{0}$ the threshold parameter. On the OPE side, we have
 \begin{equation}
 \Pi_{1}^{OPE}=\int_{0}^{\infty} ds \frac{\frac{-1}{3 \pi^{2}} \langle\overline{u} u\rangle s}{s-q^2} +\frac{1}{24\pi^{2}}\ln(-q^{2})g_{s}\langle\overline{u}\sigma_{\lambda\sigma}\frac{\lambda^{a}}{2}u G_{\lambda\sigma}^{a}\rangle,
 \end{equation}
 \begin{equation}
 \Pi_{2}^{OPE}=\frac{1}{\pi}\int_{0}^{\infty} ds \frac{-\frac{1}{20 (2\pi)^{3}} s^{2}-\frac{14}{15(2\pi)^{3}}\pi^{4} T^{4}}{s-q^{2}}+\frac{4}{3}\langle\overline{u}u\rangle^{2}\frac{1}{q^{2}}.
 \end{equation}
 \par Matching both representations and taking Borel transformation, it is obtained
 \begin{equation}\label{sum:origin1}
 \lambda_{\Delta}^{2}M_{\Delta}e^{-M_{\Delta}^{2}/M^{2}}=\frac{-\langle\overline{u}u\rangle}{3\pi^{2}}
                                                                 M^{4}[1-(1+\frac{s_{0}}{M^{2}})e^{-s_{0}/M^{2}}]
                                                                 +\frac{1}{24\pi^{2}}M^{2}M_{0}^{2}\langle\overline{u}u\rangle,
 \end{equation}
 \begin{equation}\label{sum:origin2}
 \lambda_{\Delta}^{2} e^{-M_{\Delta}^{2}/M^{2}}=\frac{1}{10(2\pi)^{4}} M^{6} [2-(2+\frac{2 s_{0}}{M^{2}}
                                                      +\frac{s_{0}^{2}}{M^{4}})e^{-s_{0}/M^{2}}]
                                                      +\frac{7}{60}T^{4}M^{2}(1-e^{-s_{0}/M^{2}})
                                                      +\frac{4}{3}\langle\overline{u}u\rangle^{2},
 \end{equation}
 with the Borel parameter $M^2$. In the calculation, we have used the parameterization for the mixed condensate\cite{art16},
 \begin{equation}
 g_{s}\langle\overline{u}\sigma_{\lambda\sigma}\frac{\lambda^{a}}{2}u G_{\lambda\sigma}^{a}\rangle=M_{0}^{2}\langle\overline{u}u\rangle,
 \end{equation}
 with $M_{0}^{2}=0.8\pm0.2GeV^{2}$.
 \par Equations (\ref{sum:origin1}) and (\ref{sum:origin2}) lead to the mass sum rule for the $\Delta$ baryon
 \begin{equation}\label{sum rule1}
 M_{\Delta}^{1}=\frac{\frac{160\pi^2}{3}abM^4-\frac{20\pi^2}{3}M_0^2aM^2}{cM^6+\frac{56}{3}\pi^4T^4dM^2+\frac{640\pi^4}{3}a^2},
 \end{equation}
 where $a=-\langle\overline{u}u\rangle$, $b=1-(1+\frac{s_0}{M^2})e^{-s_0/M^2}$, $c=2-(2+\frac{2 s_{0}}{M^{2}}+\frac{s_{0}^{2}}{M^{4}})e^{-s_{0}/M^{2}}$, and $d=1-e^{-s_{0}/M^{2}}$.
\par Another way to get the mass sum rule is from Eq. (\ref{sum:origin2}) and its derivative with respect to $\frac{1}{M^2}$, which leads to another sum rule for the $\Delta$ mass:
\begin{equation}\label{sum rule2}
M_\Delta^{2}=(\frac{eM^8+\frac{56\pi^4}{3}T^4bM^4}{cM^6+\frac{56}{3}\pi^4T^4dM^2+\frac{640\pi^4}{3}a^2})^{1/2},
\end{equation}
with $e=6-(6+\frac{6s_0}{M^2}+\frac{3s_0^2}{M^4}+\frac{s_0^3}{M^6})e^{-s_0/M^2}$.
\par Similarly, the sum rules for the masses of the $\Sigma^{*}$, $\Xi^{*}$ and $\Omega$ baryons can be obtained in the same process, taking advantage of the results in Ref. \cite{art17} for the condensates parts. The results are, to order $m$,
\begin{equation}
M_{\Sigma^{*}}^{1}=\frac{\frac{160\pi^2}{9}(2a+f)bM^4-\frac{20\pi^2}{9}M_0^2(2a+f)M^2+5mcM^6+\frac{320\pi^2}{3}ma^2+\frac{56\pi^4}{3}mT^4dM^2}{cM^6+\frac{56}{3}\pi^4T^4dM^2+\frac{40\pi^2}{3}m(4a-f)M^2+\frac{640\pi^4}{9}(a^2+2af)},
\end{equation}
\begin{equation}
M_{\Sigma^{*}}^{2}=(\frac{eM^8+\frac{56\pi^4}{3}T^4bM^4-\frac{40\pi^2}{3}m(4a-f)M^4}{cM^6+\frac{56}{3}\pi^4T^4dM^2+\frac{40\pi^2}{3}m(4a-f)M^2+\frac{640\pi^4}{9}(a^2+2af)})^{1/2},
\end{equation}
\begin{equation}
M_{\Xi^{*}}^{1}=\frac{\frac{160\pi^2}{9}(a+2f)bM^4-\frac{20\pi^2}{9}M_0^2(a+2f)M^2+10mcM^6+\frac{640\pi^2}{9}m(2a^2+af)+\frac{112\pi^4}{3}mT^4dM^2}{cM^6+\frac{56}{3}\pi^4T^4dM^2+\frac{80\pi^2}{3}m(2a+f)M^2+\frac{640\pi^4}{9}(b^2+2af)},
\end{equation}
\begin{equation}
M_{\Xi^{*}}^{2}=(\frac{eM^8+\frac{56\pi^4}{3}T^4bM^4-\frac{80\pi^2}{3}m(2a+f)M^4}{cM^6+\frac{56}{3}\pi^4T^4dM^2+\frac{80\pi^2}{3}m(2a+f)M^2+\frac{640\pi^4}{9}(b^2+2af)})^{1/2},
\end{equation}
\begin{equation}
M_{\Omega}^{1}=\frac{\frac{160\pi^2}{3}fbM^4-\frac{20\pi^2}{3}M_0^2fM^2+15mcM^6+320\pi^4mf^2+56\pi^4mT^4dM^2}{cM^6+\frac{56}{3}\pi^4T^4dM^2+120\pi^2mfM^2+\frac{640\pi^4}{3}f^2},
\end{equation}
\begin{equation}
M_{\Omega}^{2}=(\frac{eM^8+\frac{56}{3}\pi^4T^4bM^4-120\pi^2mfM^4}{cM^6+\frac{56}{3}\pi^4T^4dM^2+120\pi^2mfM^2+\frac{640\pi^4}{3}f^2})^{1/2},
\end{equation}
where $f=-\langle\overline{s}s\rangle$ and $m$ is the mass of the $s$ quark.
\par In order to do numerical analysis, we have to know the relations between thermal expectation of operators in our OPE computation and the corresponding vacuum condensate. To first order in the pion distribution, the thermal average of an operator $O$ is given by \cite{art18}:
\begin{equation}
 \langle O\rangle=\langle0|O|0\rangle+\sum_{i=1,2,3}\int\frac{d^3k n_1(\omega_1)}{(2\pi)^32\omega_1}\langle\pi^i(x)|O|\pi^i(x)\rangle.
 \end{equation}
 \par Making use of the soft $\pi$-meson methods, the pion matrix element can be reduced to the vacuum expectation value of a double commutator,
 \begin{equation}
 \langle O\rangle=\langle0|O|0\rangle-\frac{1}{F_\pi^2}\sum_{i=1,2,3}\int\frac{d^3k n_1(\omega_1)}{(2\pi)^32\omega_1}\langle0|[Q_5^i,[Q_5^i,O]]|0\rangle,
\end{equation}
 where $Q_5^i=\int d^3x A_0^i(x)$ is the axial-vector charge. For light quark case, the result is,
\begin{equation}
\langle\overline{q}q\rangle=\langle0|\overline{q}q|0\rangle (1-\frac{T^{2}}{8F_{\pi}^{2}}),
\end{equation}
with $F_{\pi}=0.092\mbox{GeV}$ being the decay constant of $\pi$ meson.
\section{Numerical analysis and Summary}\label{sec3}

Before the numerical analysis of the mass QCD sum rules, we first need to know the input parameters of the QCD vacuum condensates. The relation between the condensate of $u$ or $d$ quark and the condensate of $s$ quark is $\langle0|\overline{s}s|0\rangle= 0.8\langle0|\overline{q}q|0\rangle$ \cite{art19}. Moreover, we use QCD inputs: $M_{0}^{2}=0.8\mbox{GeV}^2$ in Ref. \cite{art19}, $\langle0|\overline{u}u|0\rangle=-0.014\mbox{GeV}^3$ in Ref. \cite{art20} and $m=0.14\mbox{GeV}$ in Ref. \cite{art21}, taking into account the effect of renormalization.
\par The sum rules also contain two auxiliary parameters: the Borel parameter $M^2$ and the continuum threshold $s_{0}$. These are not physical quantities, hence the physical observable should be approximately insensitive to them. Therefore, we look for working regions of these parameters such that the dependence of the mass on these parameters are weak. Generally, the continuum threshold $s_{0}$ is related to the square of the first exited state which has the same quantum numbers as the concerned hadron, while the Borel parameter $M^2$ is determined by demanding that both the contributions of the higher states and continuum are sufficiently suppressed and the contributions coming from higher dimensional operators have a good convergence.
\par
\begin{figure}
\subfigure[]{
\epsfxsize=7cm{\epsffile{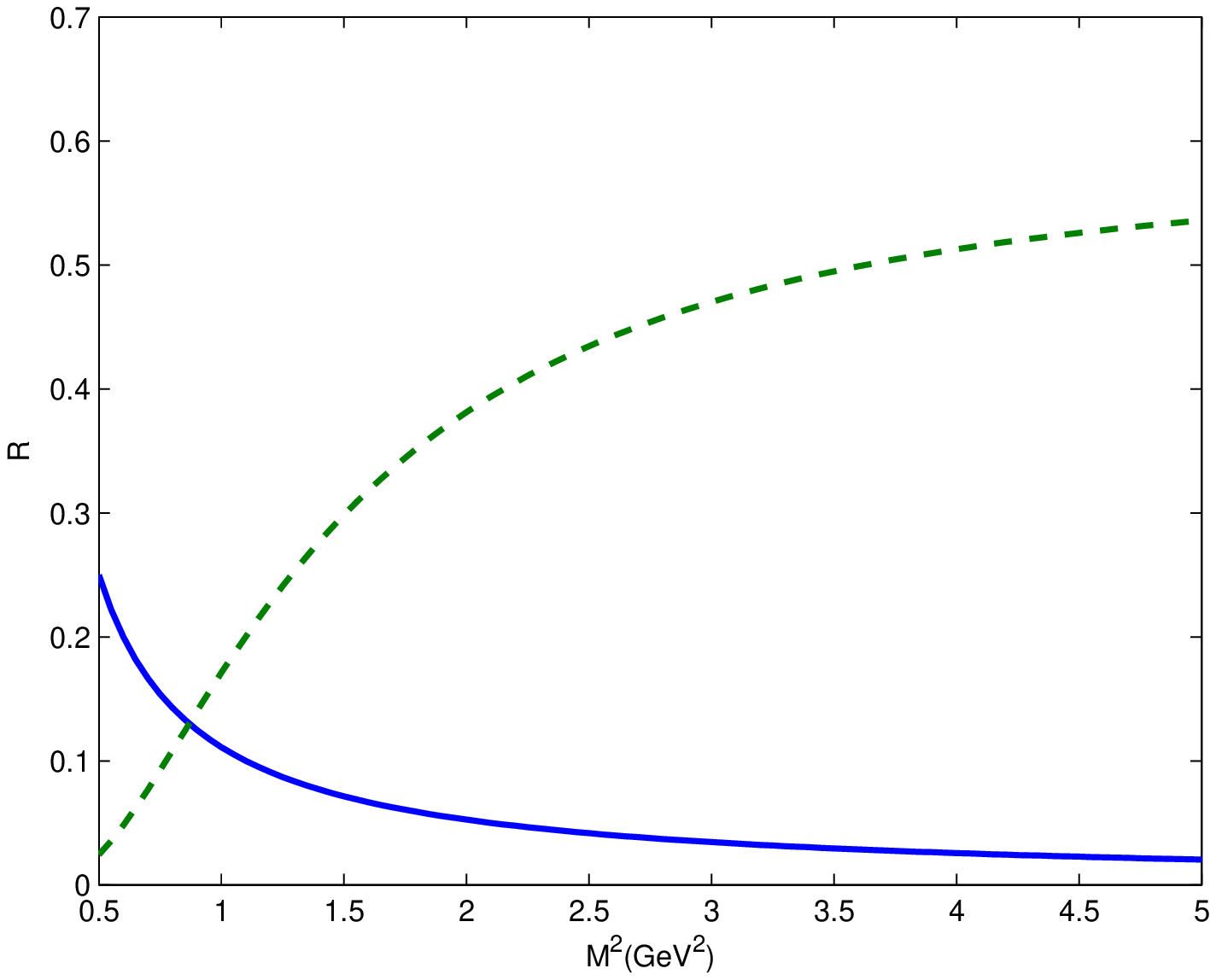}}}
\subfigure[]{
\epsfxsize=7cm{\epsffile{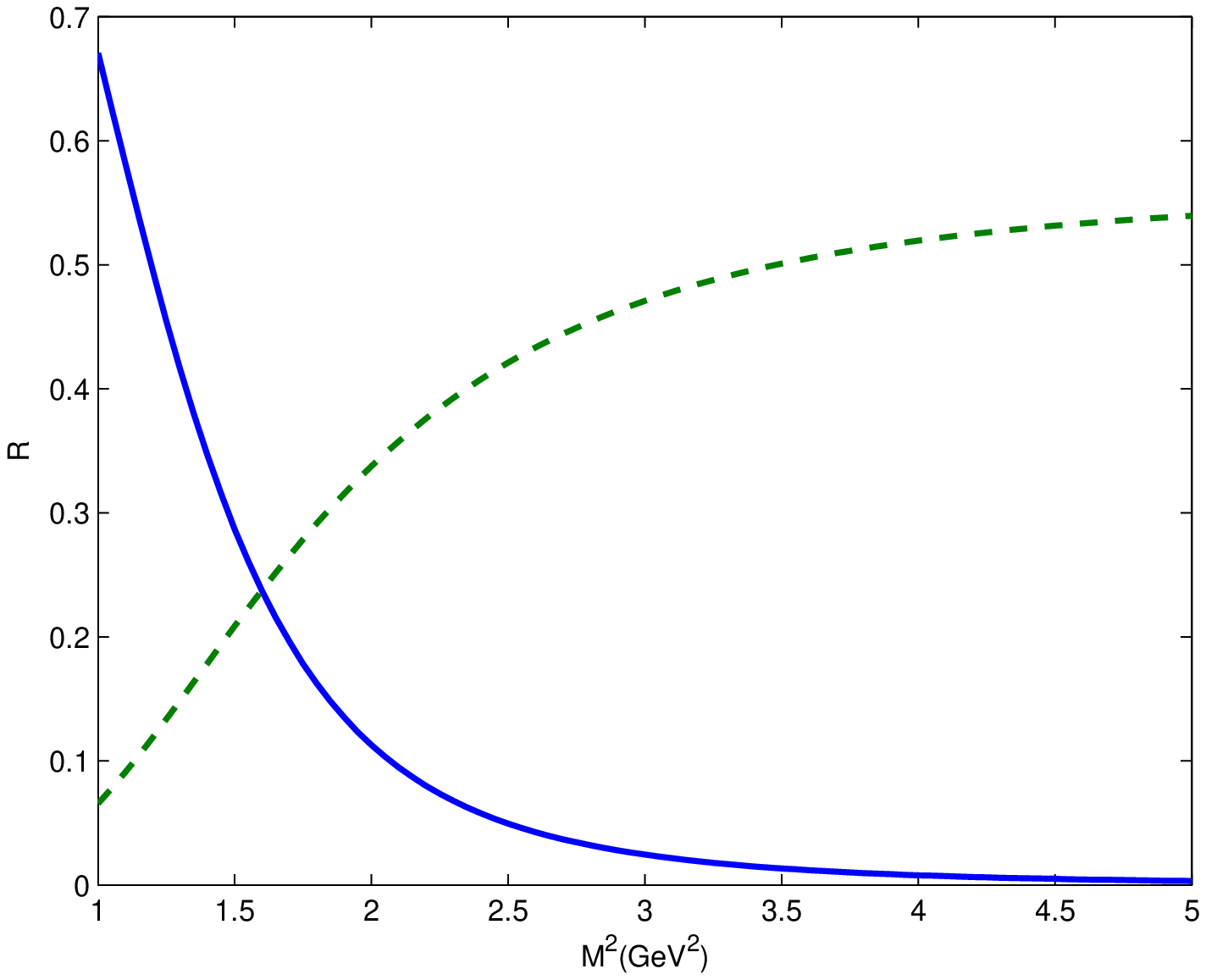}}}
\caption{The solid line represents the ratio of the contribution of the highest-dimension operator to the total contribution; the dashed line stands for the ratio of the contribution of the excited and continuum states to the total one. Figs. (a) and (b) represent the $g_{\mu\nu}$ and the $g_{\mu\nu}\not\!{q}$ structures, respectively.}\label{fig1}
\end{figure}

\begin{figure}
\subfigure[]{
\epsfxsize=7cm{\epsffile{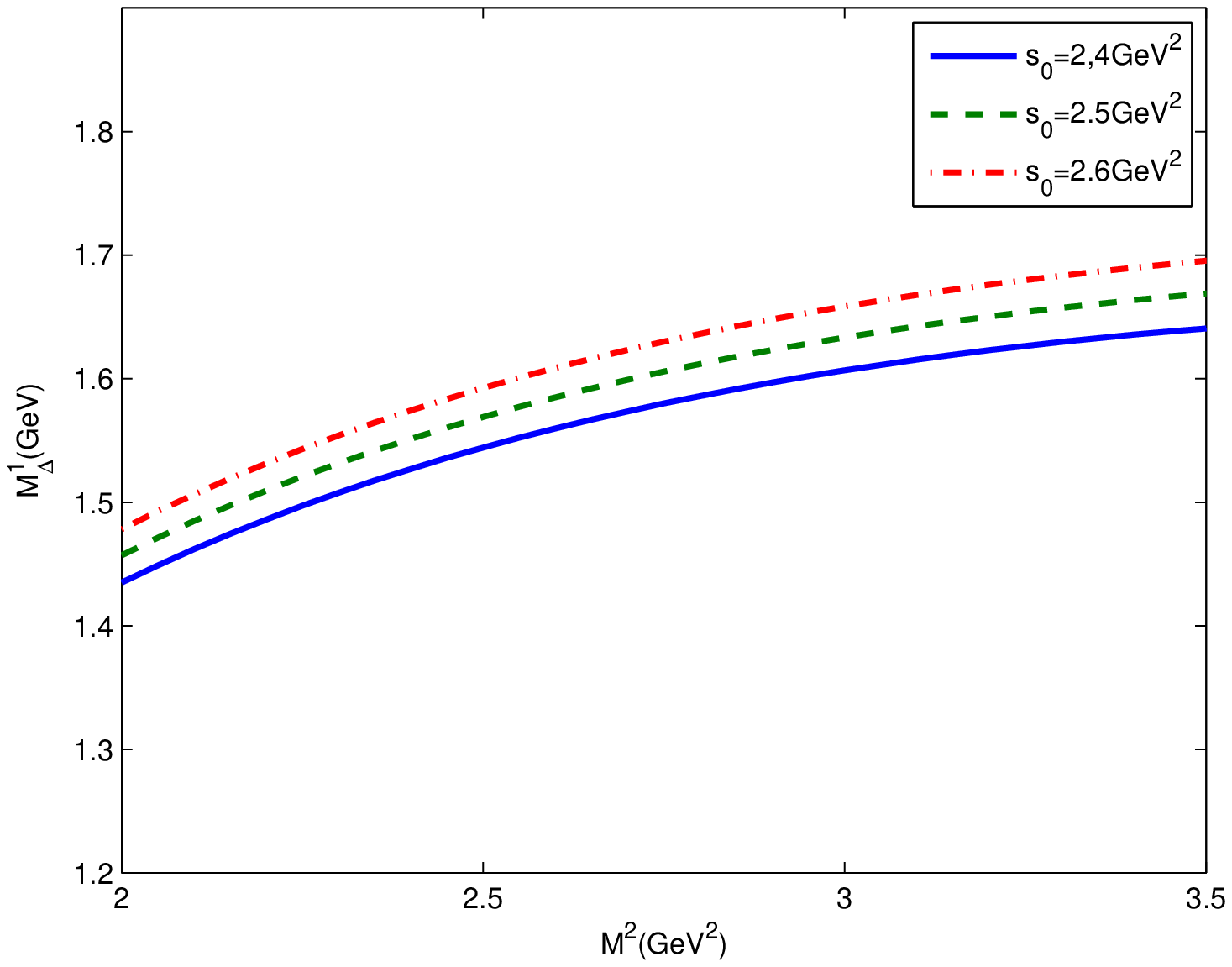}}}
\subfigure[]{
\epsfxsize=7cm{\epsffile{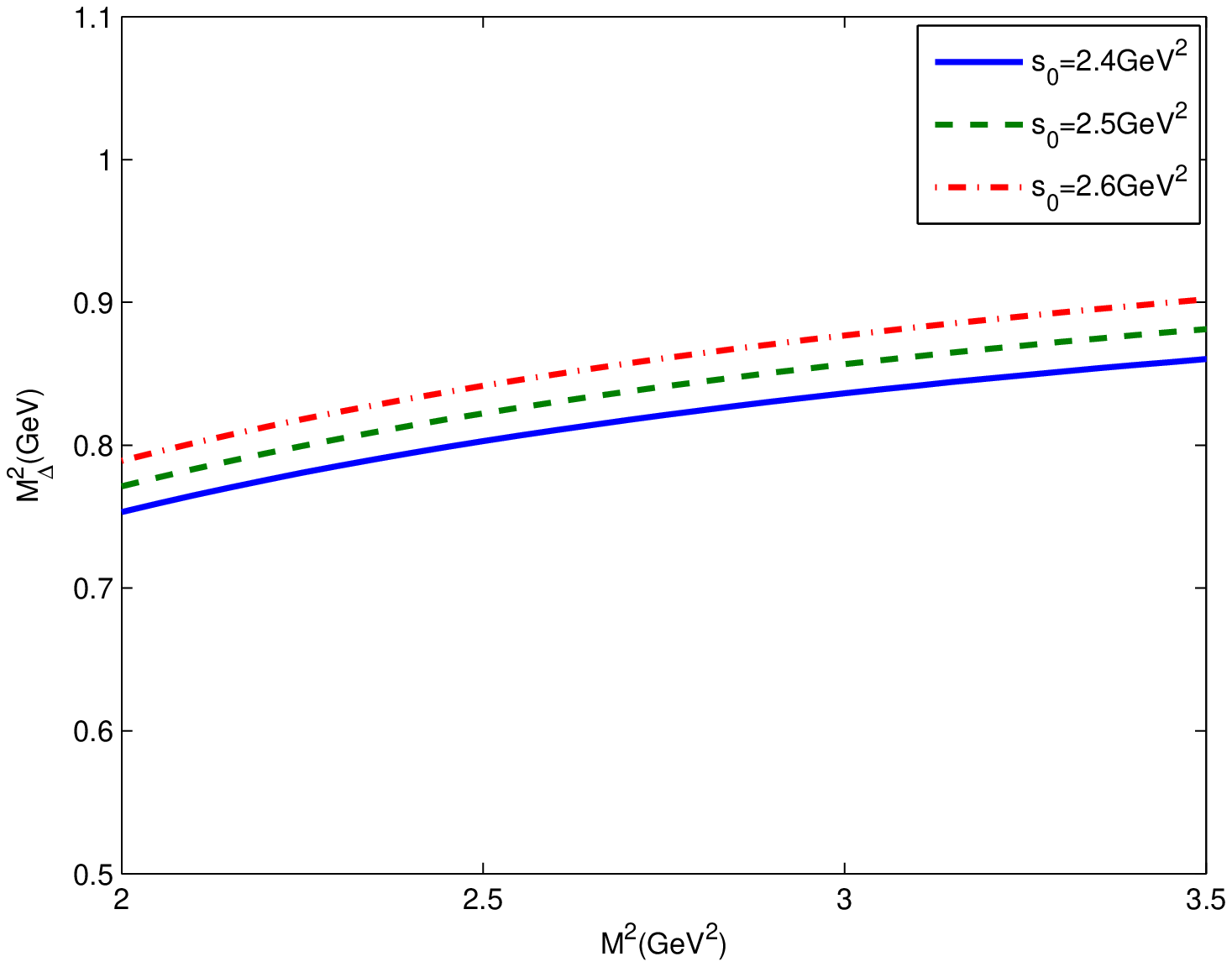}}}
\subfigure[]{
\epsfxsize=7cm{\epsffile{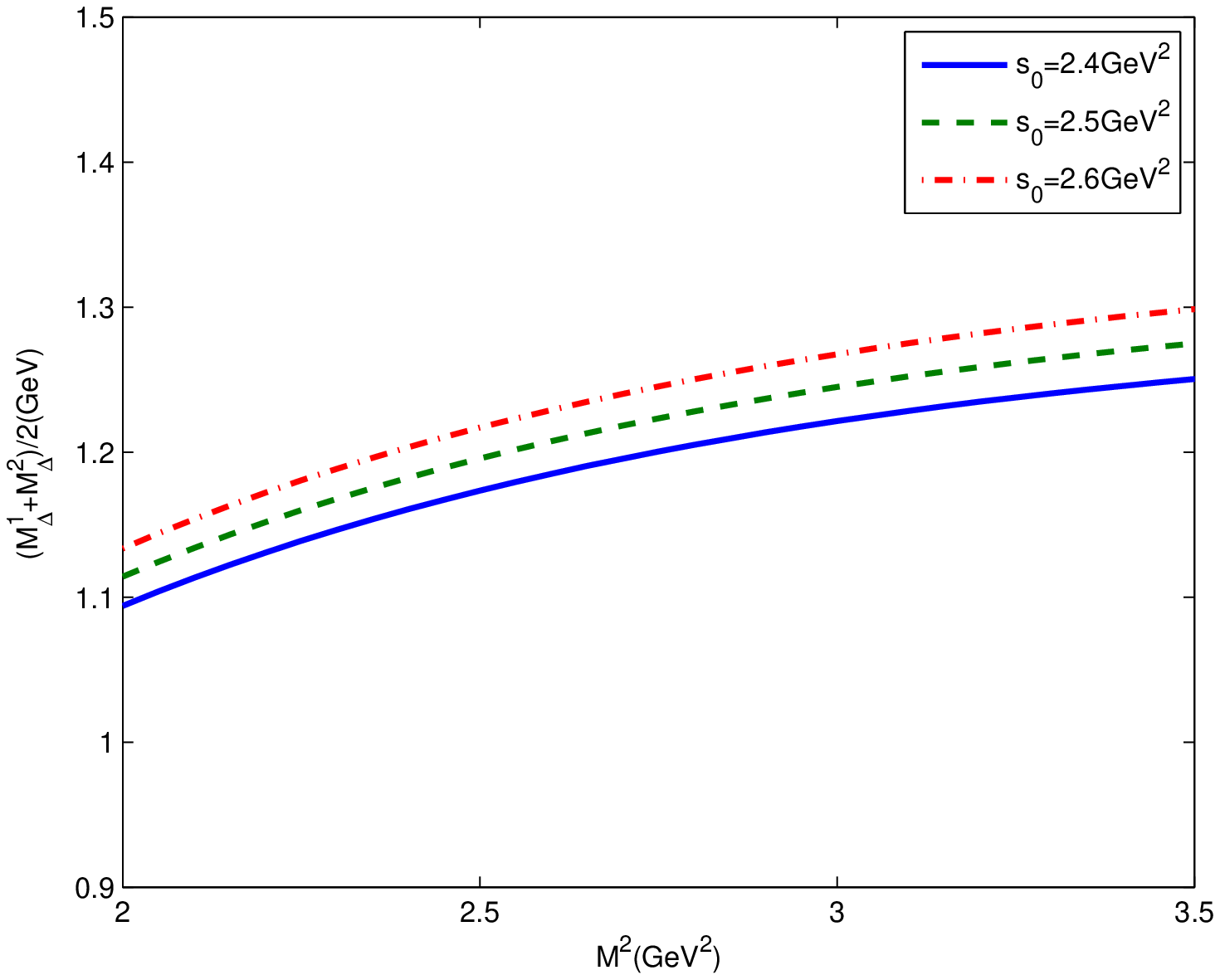}}}
\caption{(a) The sum rule (\ref{sum rule1}) at $T=0GeV$. (b) The sum rule (\ref{sum rule2}) at $T=0GeV$. (c) The average sum rule$\frac{M_\Delta^1+M_\Delta^2}{2}$ at $T=0GeV$.}\label{fig2}
\end{figure}

 In Fig.\ref{fig1} we show the contributions from the excited and continuum states and from the higher-dimension operators in the OPE computation for the $\Delta$ case. The lower limit of the Borel parameter $M^2$ is determined by demanding that the contribution of the higher-dimension operator is less than $10\%$ of the total contribution and the upper value of $M^2$ is the point at which the contribution of the excited and continuum states is $50\%$ of the total one. The working window is $2\mbox{GeV}^2\leq M^2 \leq 3.5\mbox{GeV}^2$ as we can see from Fig.\ref{fig1}. In addition, the physical quantities should vary weakly with the Borel parameter. Thus the proper range of $M^2$ in our case is chosen as $2.6\mbox{GeV}^2\leq M^2 \leq 3.5\mbox{GeV}^2$. For the parameter $s_0$, we take values in some range of the square of the first excited state of the baryon $\Delta$, $s_0=2.4 \mbox{GeV}^2, s_0=2.5 \mbox{GeV}^2, s_0=2.6 \mbox{GeV}^2$ in our case, and find that its effect can be neglected within the current accuracy. Because we don't know which sum rule is better than the other, we analyze the mass sum rules (\ref{sum rule1}), (\ref{sum rule2}) and their average at $T=0\mbox{GeV}$ in Fig.\ref{fig2}, finding that the average sum rule works well which gives the mass estimation $M_\Delta=1.20\pm 0.10\mbox{GeV}$.
\par
\begin{figure}
\subfigure[]{
\epsfxsize=7cm{\epsffile{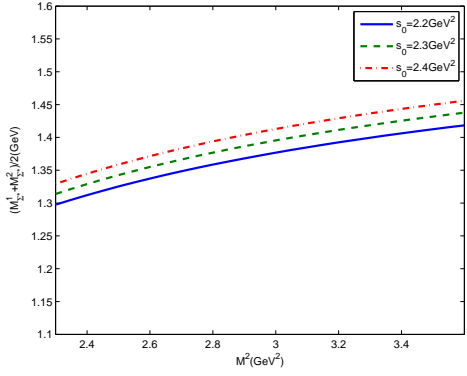}}}
\subfigure[]{
\epsfxsize=7cm{\epsffile{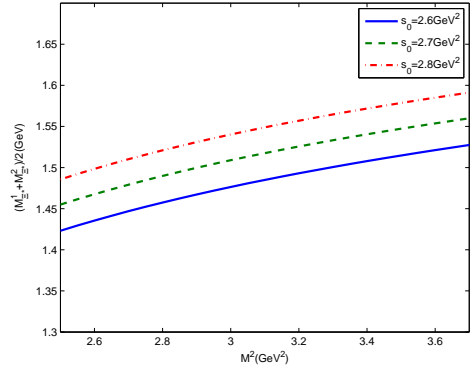}}}
\subfigure[]{
\epsfxsize=7cm{\epsffile{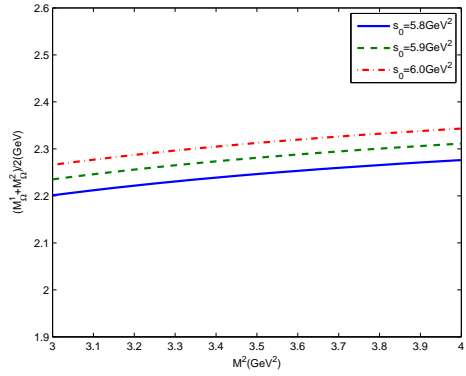}}}
\caption{(a) The average sum rule of $\Sigma^*$ mass at $T=0GeV$. (b) The average sum rule of $\Xi^*$ mass at $T=0GeV$. (c) The average sum rule of $\Omega$ mass at $T=0GeV$. }\label{fig3}
\end{figure}

 The same analysis are carried out for the cases of the other three baryons. With the same criterion of determining the lower and upper values of the Borel parameter $M^2$, the results are the following: $2.3\mbox{GeV}^2\leq M^2 \leq 3.5\mbox{GeV}^2$ for the $\Sigma^*$ baryon, $2.5\mbox{GeV}^2\leq M^2 \leq 3.7\mbox{GeV}^2$ for the $\Xi^*$ baryon, and $3\mbox{GeV}^2\leq M^2 \leq 4\mbox{GeV}^2$ for the $\Omega$ baryon. Fig. \ref{fig3} shows the average sum rules for the other three baryons at $T=0\mbox{GeV}$, from which we get the numerical estimates $M_{\Sigma^*}=1.37\pm 0.08\mbox{GeV}$, $M_{\Xi^*}=1.50\pm0.09\mbox{GeV}$, and $M_{\Omega}= 2.28\pm0.08\mbox{GeV}$. These results show that our sum rules for the masses of the decuplet baryons are meaningful and the results agree with the experimental results. In table \ref{table} we list the spectra of the decuplet baryons and the comparison with the experimental results.

\begin{table}
\caption{ The mass spectra of the decuplet baryons.}\label{table}
\begin{tabular}{|c|c|c|}
  \hline
  Baryon      &   Experimental($\mbox{Mev}$)\cite{art20}   &    This work($\mbox{GeV}$) \\
  \hline
  {$\Delta$}  &      $1232\pm3$                                &       $1.20\pm0.10$ \\
  \hline
  {$\Sigma^*$} &     $1382\pm2$                                 &       $1.37\pm0.08$ \\
  \hline
  {$\Xi^*$}    &     $1533\pm1$                                 &       $1.50\pm0.09$ \\
  \hline
  {$\Omega$}  &    $2252\pm9$                                  &       $2.28\pm0.08$ \\
  \hline
\end{tabular}
\end{table}

\par
\begin{figure}
\subfigure[]{
\epsfxsize=7cm{\epsffile{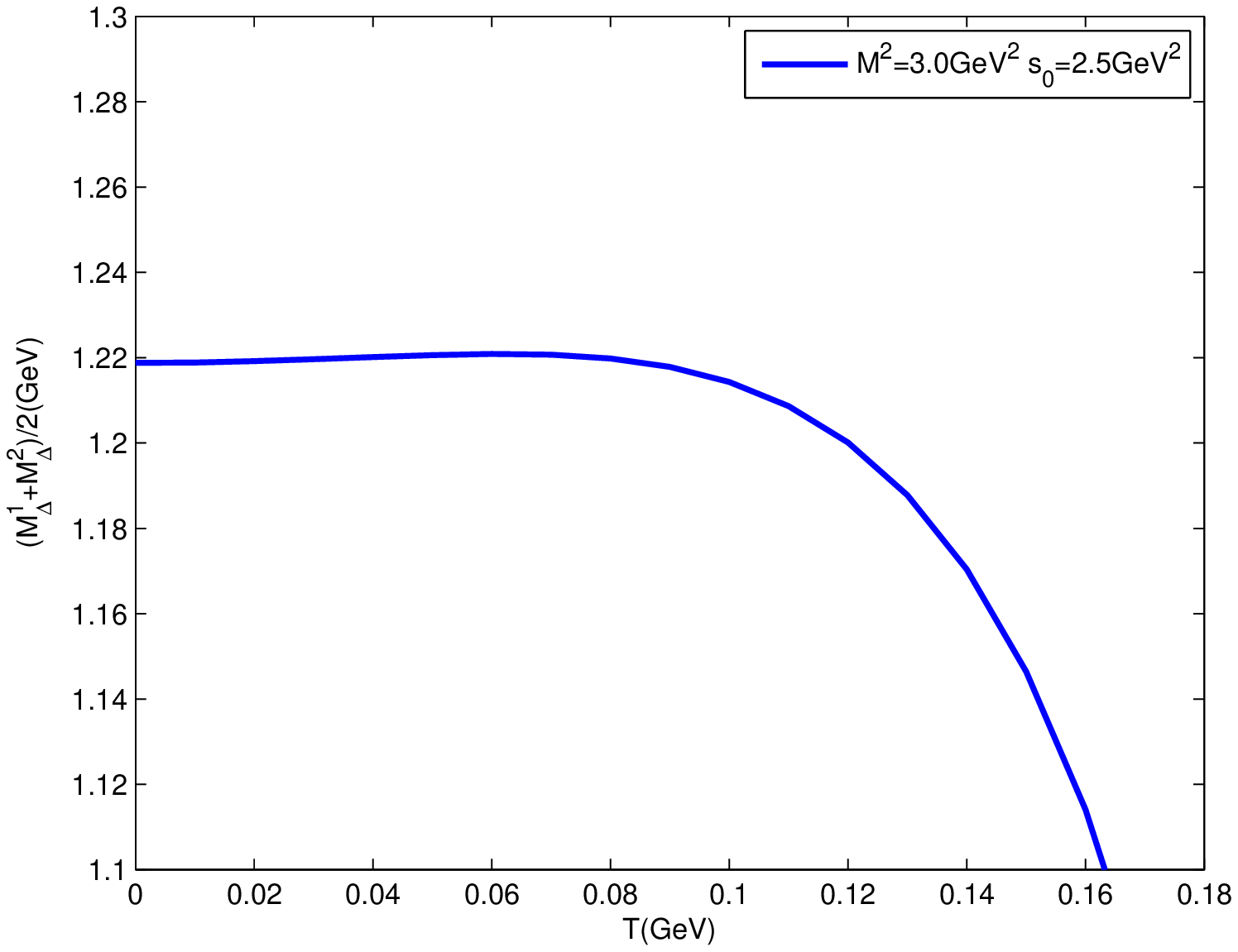}}}
\subfigure[]{
\epsfxsize=7cm{\epsffile{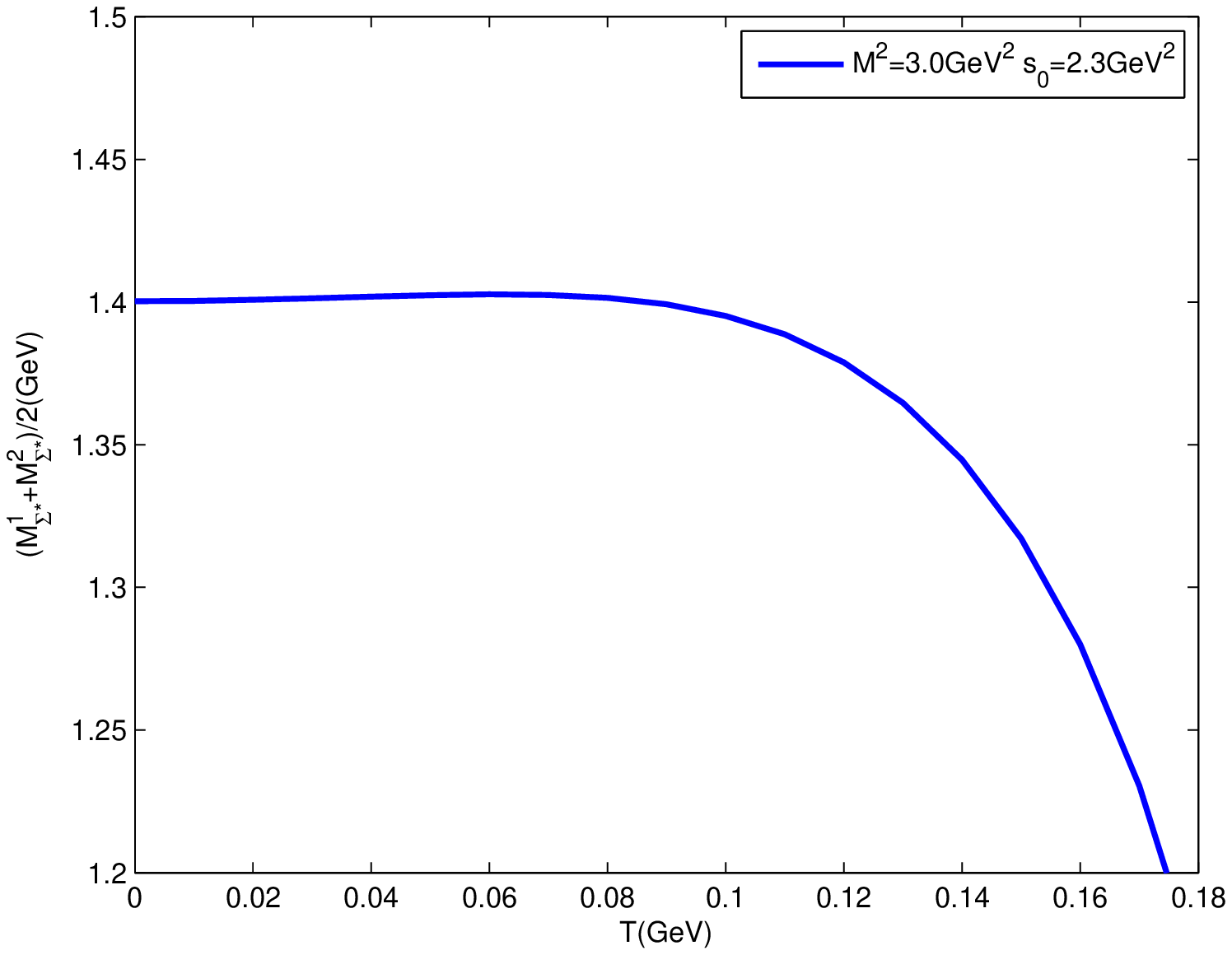}}}
\subfigure[]{
\epsfxsize=7cm{\epsffile{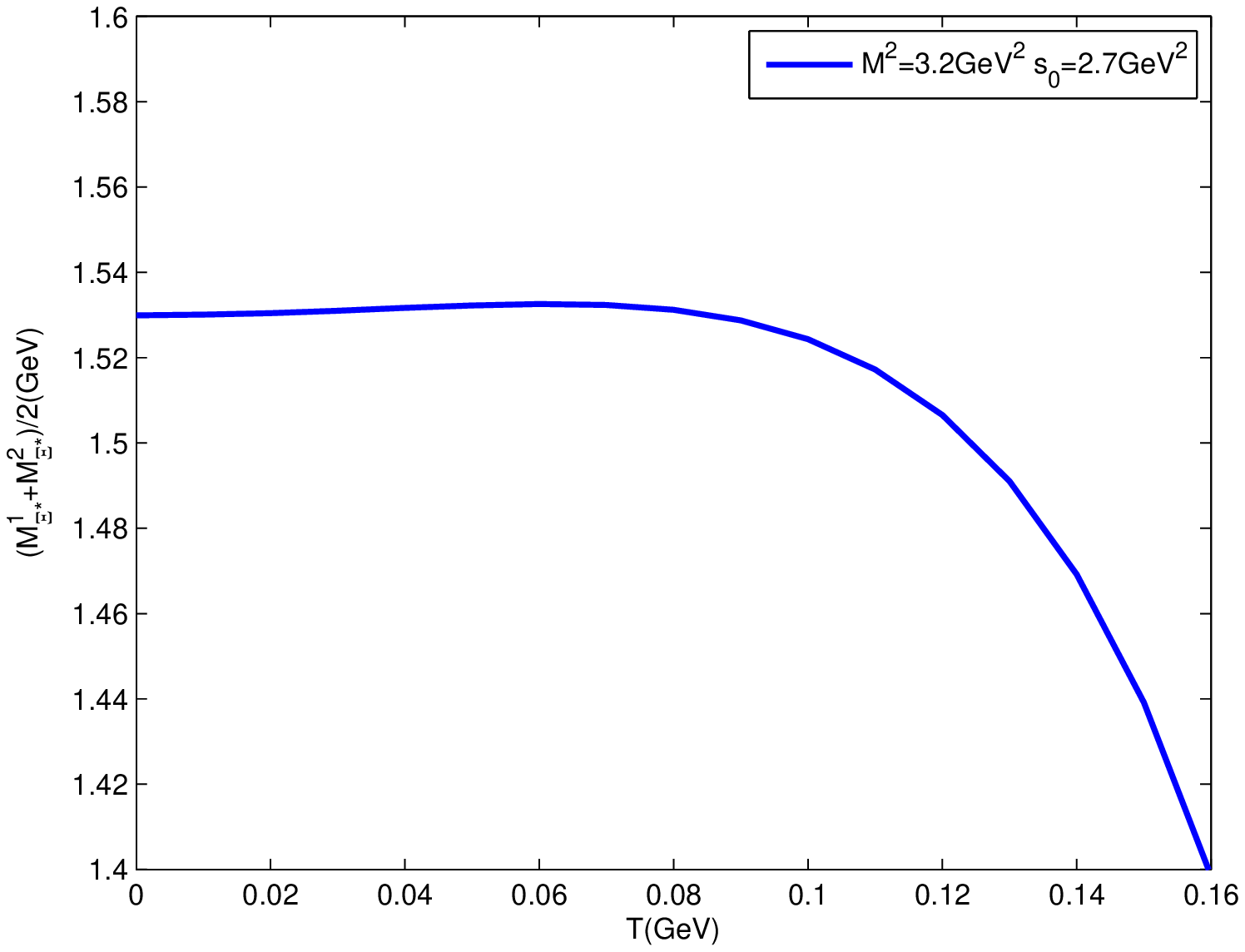}}}
\subfigure[]{
\epsfxsize=7cm{\epsffile{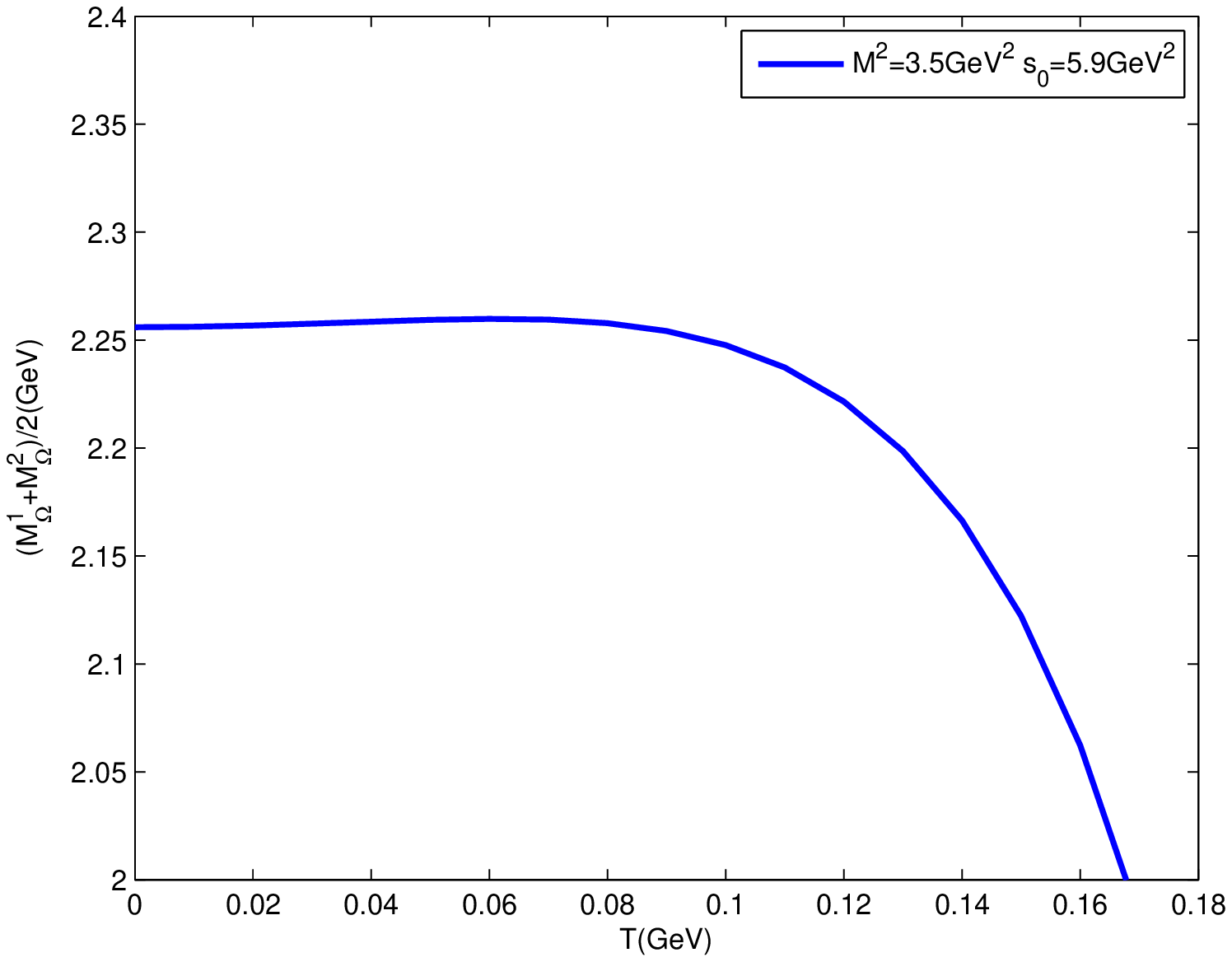}}}
\caption{(a) The variations of $\frac{M_{\Delta}^{1}+M_{\Delta}^{2}}{2}$ with temperature $T$. (b) The variations of $\frac{M_{\Sigma^*}^1+M_{\Sigma^*}^2}{2}$ with temperature $T$. (c) The variations of $\frac{M_{\Xi^*}^1+M_{\Xi^*}^2}{2}$ with temperature $T$. (d) The variations of $\frac{M_\Omega^1+M_\Omega^2}{2}$ with temperature $T$.}\label{fig4}
\end{figure}

 Fig.\ref{fig4} presents the temperature dependence of the baryons' masses. It is found that the masses nearly vary with temperature below $T \leq 0.11 \mbox{GeV}$, but decrease with increasing temperature as $T \geq 0.11 \mbox{GeV}$. Because the hadrons undertake a phase transition from hadron states to quark-gluon plasm at sufficiently high temperature, the results show that the baryons remain as hadronic states as $T\leq0.11\mbox{GeV}$ and the critical temperature of phase transition to quark-gluon plasma is above $T=0.11 \mbox{GeV}$, ie. $T_c\geq 0.11 \mbox{GeV}$. In fact, as the temperature is increased, the hadron melts and the width of the hadron increases until it becomes infinite at the critical temperature.
\par In summary, we study the temperature dependence of masses of the decuplet baryons using the thermal QCD sum rules. For the terms containing condensates, we use the results obtained by the OPE method and replace the vacuum condensates by the thermal expectation values at finite temperature. The extension of QCD sum rules to finite temperature is based on the following assumption: the operator product expansion(OPE) and notion of quark-hadron duality remain valid at finite temperature, but the vacuum condensates must be replaced by their thermal expectation values. Adopting the quark propagator in the coordinate space at finite temperature, we calculate the perturbative part of the thermal two-point correlator to $T^{8}$ order. It is found that the terms containing $T^{8}$ order have no contribution to the spectral functions and the temperature corrections mainly come from $T^4$ terms. The calculations indicate that the baryon ``melts" or the hadron-quark phase transition occurs at the temperature $T \geq 0.11\mbox{GeV}$ for the decuplet baryons.


\acknowledgments  This work was supported in part by the National
Natural Science Foundation of China under Contracts No.11475257 and No.11275268, and NUDT Foundation under Contract JC14-02-05.


\end{document}